\documentstyle[12pt]{article}
\newcommand{\dd}{{\rm d}}

\newcommand{\sig}[2]{\sigma^{#1}_{\,\,#2}}

\def\ep{\eta}

\def\be{\begin{equation}}
\def\ee{\end{equation}}
\def\bea{\begin{eqnarray}}
\def\eea{\end{eqnarray}}
\def\l{\label}
\def\c{\cite}

\def\hsp5{\hspace{5mm}}
\newcommand{\sfrac}[2]{{\textstyle{#1\over#2}}}
\def\case#1/#2{\textstyle\frac{#1}{#2}}
\def\l{\label}
\def\ct{\cite}

\def\Th{\Theta}
\def\sig{\sigma}
\def\om{\omega}
\def\udot{\dot{u}}
\def\3nab{\tilde{\nabla}}

\def\lgl{\langle}
\def\rgl{\rangle}

\def\c{\mbox{curl}}
\def\div{\mbox{div}}

\def\hsp5{\hspace{5mm}}
\def\case#1/#2{\textstyle\frac{#1}{#2}}

\def\be {\begin{equation}}
\def\ee {\end{equation}}
\def\bea {\begin{eqnarray}}
\def\eea {\end{eqnarray}}

\def\case#1/#2{\textstyle\frac{#1}{#2} }

\title{Shear free solutions in General Relativity Theory}
\author{George F R Ellis, \\ Mathematics Department, University of
Cape Town\footnote{email: george.ellis@uct.ac.za}}
\begin{document}
\maketitle

\begin{abstract}
The Goldberg-Sachs theorem is an exact result on shear-free null geodesics in a
vacuum spacetime. It is compared and contrasted with an exact result for
pressure-free matter: shear-free flows cannot both expand and rotate. In
both cases, the shear-free condition restricts the way distant matter can
influence the local gravitational field. This leads to intriguing discontinuities
in the relation of the General Relativity solutions to Newtonian solutions in the
timelike case, and of the full theory to the linearised theory in the null case.\\

It is a pleasure to dedicate this paper to Josh Goldberg.\\

\end{abstract}


\section{Introduction}
The physical interpretation of solutions of the Einstein Field
Equations in General Relativity Theory is intimately tied in to the
way timelike and null geodesics behave. The differential properties
of families of geodesics are described by their expansion, rotation
and shear in the timelike case \cite{Ehl61,Ell71}, and by the null
expansion and shear in the case of the irrotational null geodesic
congruences that underlie observations \cite{EhlSac61}. This paper
discusses the key role of shear in physical processes as evidenced
by their effect on such congruences, and hence the very special
nature of shear-free solutions. The remarkable Goldberg-Sachs
theorem \cite{GolSac62} demonstrates this very special nature in the
case of shear free null geodesics. It was proceeded by G\"{o}del's
intriguing results shear-free timelike geodesics
 \cite{God52}, which considered specific the case of spatially
homogeneous geometries. This result was generalized to the
inhomogeneous case of any shearfree timelike geodesics in
\cite{Ell67}. Although very different in detail, the timelike and
null cases are in a sense analogous results: they both refer to the
way that shear in a congruence conveys information about distant
matter, so shearfree congruences can only occur in restricted
circumstances. That is what is explored in this paper.

The Einstein Field Equations (`EFE') take the form
\begin{equation}  \label{eq:efe} G_{ab} \equiv R_{ab} -
\frac{1}{2}\,R\,g_{ab} = \kappa T_{ab} - \Lambda\,g_{ab} \ ,
\end{equation}
showing how matter causes space time curvature by specifying the
spacetime Ricci tensor $R_{ab}$ in terms of the matter stress tensor
$T_{ab}$. Provided the cosmological constant $\Lambda$ is constant
in time and space, the twice-contracted Bianchi identities together
with (\ref{eq:efe}) guarantee the conservation of total
energy-momentum:
\begin{equation}  \label{eq:cons}
\{\nabla_{b}G^{ab} = 0, \,\,\, \nabla_{a}\Lambda = 0\} \hsp5
\Leftrightarrow \hsp5  \nabla_{b}T^{ab} = 0.
\end{equation}
To complete the dynamical description, we must specify the matter
present by providing suitable equations of state relating the
components of $T_{ab}$.

The locally free gravitational field is given by Weyl tensor
$C_{abcd}$, which is the trace-free part of the full curvature
tensor $R_{abcd}$. It is the part of the spacetime curvature not
directly determined pointwise by matter but rather determined by
matter elsewhere through tidal effects and gravitational waves. The
Bianchi identities
\begin{eqnarray}
\nabla_{[e}R_{ab]cd}=0 \,\, \Leftrightarrow \,\, \nabla^ d C_{ a b c
d } =- \nabla_{[a}\Big\{ R_{b] c} - {1\over6}Rg_{b] c} \Big\},
\label{rbi}
\end{eqnarray}
are integrability conditions relating the Ricci tensor to the Weyl
tensor \cite{Ell71,EllVan99,TsaChaMaa08}, mediating the action at a
distance of the gravitational field; they take a form similar to
Maxwell's equations for the electromagnetic field \cite{MaaBas98}.
Together with suitable equations of state for the matter, relating
the various components of $T_{ab}$, equations (\ref{eq:efe}),
(\ref{eq:cons}), (\ref{rbi}) determine the dynamical evolution of
the model and the matter in it.

I will simplify things in what follows by only considering the
simplest form of matter: a pressure-free perfect fluid, such as cold
dark matter or cold baryonic matter. This moves with a unique
4-velocity $u^a = dx^a/ds$ where $s$ is proper time along the matter
flow lines, so $u^au_a = -1$. The matter energy-momentum tensor
$T_{ab}$ then takes the form
 \bea T_{ab} = \rho\,u_a\,u_b, \,\,\rho \geq 0\,,
\l{eq:stress} \eea where $\rho = T_{ab}u^{a}u^{b}$ is the energy
density. The energy-momentum conservation equations (\ref{eq:cons})
reduce to
\be \l{eq:en} \dot{\rho} = -\rho\,\nabla^a u_a \,,\,\, \udot_{a} :=
u^b\nabla_b u^a =0  \,,\ee so the matter, affected only by gravity
and inertia, moves geodesically. This is the case of pure
gravitation: it separates out the (non-linear) gravitational effects
from the complexities of thermodynamic and fluid dynamical effects.


\section{Timelike flows}
For a given fundamental observer moving with 4-velocity $u^a$,
spacetime decomposes into space and time \cite{Ehl61,Ell71}. The
metric of the tangent spaces orthogonal to $u^a$ is given by
\begin{equation}
h_{ab}=g_{ab}+ u_au_b\,\Rightarrow\,\,\,h^a_ch^c_b=h_b^a,\,\,\,
h^a{}_a=3, \,\,\, h_{ab}u^b=0
\end{equation}
where $g_{ab}$ is the spacetime metric. The metric $h_{ab}$ is used
to project orthogonally to $u^a$; the volume element for tangent
3-spaces orthogonal to $u^a$ is $\eta_{abc} = \eta_{abcd}u^d =
\eta_{[abc]}$. Here round brackets denote symmetrization, square
brackets denote skew symmetrization, and angle brackets represent
the Projected Symmetric Trace-Free (`PSTF') part of a tensor. The
covariant derivative of a geodesic timelike vector field $u^a$ may
be split into irreducible parts as
\begin{equation}\label{Decomposeu}
\nabla_bu_a= {1\over3}\Theta h_{ab}+\ep_{abc}\omega^c
+\sigma_{ab}\,,\,\,\, \sigma_{ab} = \sigma_{<ab>}, \,\,\,
\end{equation}
where
\begin{eqnarray}
\Theta\equiv h^{ab}\nabla_{a} u_{b} \mbox{ , } \,\,\,\,
\sigma_{ab}\equiv
h^{c}_{a}h^{d}_{b}\nabla_{(c}u_{d)}-\frac{1}{3}\Theta h_{ab} \mbox{
,} \,\,\,\,\,\omega^a=\frac{1}{2}\eta^{abc}\nabla_{[c}u_{d]}
\end{eqnarray}
are the rate of expansion, shear, and vorticity respectively. Time
and spatial derivatives relative to $u^a$ for a tensor $T^{a}{}_{b}$
are defined by:
\begin{equation}
\dot{T}^{a}{}_{b}= u^c \nabla_c T^{a}{}_{b}\,,~ \nabla_c T^{a}{}_{b}
= h_c{}^d h^a{}_e h_b{}^f \nabla_d T^{e}{}_{f}\,.
\end{equation}
 The Weyl tensor splits into the PSTF gravito-electric and gravito-magnetic fields
\begin{eqnarray}
E_{a b }=C_{a  c b  d}u^{ c}u^ d\,,~~ H_{ a b }={{1\over2}}\ep_{a c
d }C^{ c  d }{}{}_{b  e}u^ e  \,,
\end{eqnarray}
which provide a covariant description of tidal forces and
gravitational radiation.

The Einstein equation (\ref{eq:efe}) with matter source term
(\ref{eq:stress}) and the Ricci identity
\begin{eqnarray}
\nabla_{[a} \nabla_{b]} u_ c= R_{a b  c   d}u^{ d}
\end{eqnarray}
for $u^a$ give the following evolution equations:
\begin{eqnarray}
 \dot{\Theta} + \sfrac13 \Theta^2 +\sigma_{a b }\sigma^{a b }-2\omega_{a}\omega^{a} +\sfrac12 \kappa \rho-\Lambda
 &=& 0\,, \label{e2}\\
 \dot{\sigma}_{\langle a b \rangle } +\sfrac23 \Theta\sigma_{a b } +\sigma_{ c\langle a}\sigma_{b\rangle }{}^{ c}
 +\omega_{\langle a}\omega_{b\rangle }
+ E_{a b } &=& 0\,,\label{e5}\\
 \dot{\omega}_{\langle a\rangle } +\sfrac23 \Theta\omega_{a}
 - \sigma_{ab }\omega^{b} &=& 0 \,.\label{e4}
\end{eqnarray}
Constraint equations are the identity $\3nab_{a}\om^{a} =0$, the
field equation \be \l{eq:onu} \sfrac{2}{3}\,\3nab^{a}\Th -
\3nab_{b}\sig^{ab} -(\c \,\om)^{a}   =0, \ee where the `curl' is
$(\c \,\om)^{a}=\eta^{abc} \3nab_{b}\om_c$, and an equation for the
magnetic part of the Weyl tensor: \be \l{hconstr}
 H^{ab} =  -\3nab^{\lgl a}\om^{b\rgl} + (\c\,\sig)^{ab} \  \ee where the `curl' is
 $ (\c\,\sigma)^{ab}  =  \eta^{cd\lgl
a}\,\3nab_{c}\sigma_{d}{}^{b\rgl} \ .$ Propagation equations for the
Weyl tensor are the $\dot{E}$-equation and $\dot{H}$-equations:
\bea \dot{E}^{\lgl ab\rgl} - (\c\,H)^{ab}  & = & -
\,\sfrac{1}{2}\,\kappa\rho\,\sig^{ab} - \Th\,E^{ab} + \
3\,\sig^{\lgl a}\!_{c}\,E^{b\rgl c} + \ \eta^{cd\lgl a}
\om_{c}E_{d}{}^{b\rgl}\ \ , \l{Edot} \\ \dot{H}^{\lgl ab\rgl} +
(\c\,E)^{ab}  & = & - \,\Th\,H^{ab} + 3\,\sig^{\lgl
a}\!_{c}\,H^{b\rgl c}  + \ \eta^{cd\lgl a}
  \om_{c}\,H_{d}{}^{b\rgl}\  \ ,
\l{Hdot} \eea
where the `curls' are $ (\c\,H)^{ab}  =  \eta^{cd\lgl
a}\,\3nab_{c}H_{d}{}^{b\rgl} \ ,$ $(\c\,E)^{ab}  =  \eta^{cd\lgl
a}\,\3nab_{c}E_{d}{}^{b\rgl} .$ The constraint equations are the
$(\div\,E)$ and $(\div\,H)$-equations: \bea \l{eq:dive}
\3nab_{b}E^{ab} = \sfrac{1}{3}\,\kappa\3nab^{a}\rho  +
 3\,\om_{b}\,H^{ab} + \ \eta^{abc}\,\sig_{bd}\,H_{c}{}^{d} \ , \\
 \l{eq:divH}  \3nab_{b}H^{ab} = -\kappa\rho\,\om^{a} +
3\,\om_{b}\,E^{ab}    - \sig_{bd}\,E_{c}{}^{d} \ . \eea
These equations - the exact non-linear equations for dust filled
spacetimes - are clearly generalizations of Maxwell's equations for
the electromagnetic field \cite{MaaBas98}.

The case of a vacuum (empty spacetime) is a special case of these
equations: just set $\rho = 0$. The general form of these equations
for arbitrary matter fields (including pressure, viscosity, and heat
flux terms) is given in \cite{EllVan99}.

\subsection{Dynamic effects}
By the vorticity conservation equation (\ref{e4}), the vorticity
along each world line is affected only by the expansion and shear:
for the case of pressure free matter considered here, vorticity
cannot be created or destroyed along any world line. By the
Raychaudhuri equation (\ref{e2}), the rate of expansion is increased
by a cosmological constant but decreased by any matter present
(because we have assumed the matter energy density is positive).
This is the local Ricci effect on the fluid flow (it results from
the Ricci tensor term in the EFE (\ref{eq:efe})), which occurs in a
point-by-point manner due to the matter occurring along the
worldline. The Weyl tensor cannot directly affect the expansion
rate, but it can do so by inducing shear (via the shear propagation
equation (\ref{e5})) which then induces a deceleration (by
(\ref{e2})). This is the Weyl effect on the fluid flow; it occurs
non-locally, due to matter at a distance from the world line. If
$E_{ab} = 0$ the evolution equations along each world line
(\ref{e2}) - (\ref{e4}) become ordinary differential equations
unaffected by distant matter: there are no tidal effects or
gravitational wave effects affecting the fluid flow, and each world
line evolves on its own, unaffected by what is happening elsewhere.
This is what has been called a `silent universe'
\cite{MatPanSae93,BruMatPan95} (for a recent review, see
\cite{WylVan06}.)

How do non-local effects occur? We can think of it as happening in
two ways. First, tidal action at a distance is represented by the
the $(\div\,E)$ equation (\ref{eq:dive}) with source the spatial
gradient of the energy density (e.g. scalar perturbation modes);
this can be regarded as a vector analogue of the Newtonian Poisson
equation, where by matter elsewhere generates an $E$-field here.
Similarly the $(\div\,H)$ equation (\ref{eq:divH}) shows that fluid
vorticity elsewhere generates a $H$-field here (e.g. vector
perturbation modes).

Alternatively, we can consider the evolution equations (\ref{Edot}),
(\ref{Hdot}), which together form a hyperbolic system. They show how
gravitational radiation arises: taking the time derivative of the
$\dot{E}$-equation gives a term of the form $(\c\,H){}\dot{}\,$;
commuting the derivatives and substituting from the
$\dot{H}$-equation eliminates $H$, and results in a term in
$\ddot{E}$ and a term of the form $(\c\,\c\,E)$, which together give
the wave operator acting on $E$ \ct{Haw66,DunBasEll96}; similarly
the time derivative of the $\dot{H}$-equation gives a wave equation
for $H$. This shows how matter over there can effect matter here by
generating gravitational radiation which travels here and causes a
non-zero $E$-field here, which then affects matter here (via the
shear equation (\ref{e5}). This is compatible with the constraint
equation effects just discussed, because the constraint equations
are preserved by the time evolution equations (see \cite{vanEll99}
corrected in \cite{vanEllSch00}). As the key link between the $E$
and $H$ fields in this process is via their curls, this suggests one
can characterize the existence of gravitational radiation by the
condition \be (\c\,H)^{ab} \neq 0, \,\, (\c\,E)^{ab} \neq 0
\l{radn}\ee which of course requires that both $E$ and $H$ are
non-zero.

However there is another way a non-zero Weyl tensor can be created
where there was none before: this is locally via matter shear (see
(\ref{Edot})). This emphasizes the crucial importance of shear in
gravitational dynamics.
It is not only the link whereby information on surrounding
inhomogeneities (given us via the electric part of the Weyl tensor)
alters the fluid flow here, it is also a source of the electric part
of the Weyl tensor. If the shear is zero, this link is broken, the
way distant matter can influence us here is very limited: for by
(\ref{e5}), only the vorticity can prevent the electric part of the
Weyl tensor from being zero.  But as we see in the next section,
this does not work; distant matter is constrained to acting in an
isotropic way around our world line. This is a very special
situation.

\section{Timelike shear-free results}
When the shear is zero, the expansion is isotropic; we might expect
vorticity to tend to generate anisotropy that would break this
condition. However a non-zero Weyl tensor might balance this
tendency. Specifically, on setting $\sigma_{ab}=0$ in the above
equations,(\ref{e5}) becomes a new constraint equation, along with
the old constraint (\ref{hconstr}) determining $E$ and $H$ in terms
of $\omega$:
\begin{eqnarray}
  E_{a b } &=& -\omega_{\langle a}\omega_{b\rangle},\,\,\,
  H^{ab} =  -\3nab^{\lgl a}\om^{b\rgl} \ .
 \label{e51}
\end{eqnarray}
The task now is to take take time derivatives of these constraints
to see if the shear-free equations (obtained by setting
$\sigma_{ab}=0$ in all the above equations) are consistent for some
non-trivial special cases.

This is a major calculation: the result is not obvious. It was not
initially tackled this way. First, in a remarkable pioneering paper
presented at a world mathematics congress in 1950, Kurt G\"{o}del
examined this question in the case of spatially homogeneous Bianchi
IX cosmologies. He showed \cite{God52} that in this case, a
shear-free universe could either expand or rotate, but not both; but
he did not show how he had obtained that result. In 1957,
Sch\"{u}ucking derived the G\"{o}del result in detail \cite{Sch57}.
In 1967, I used an orthonormal tetrad formalism to show that the
restriction of spatial homogeneity was unnecessary:
\begin{quote}
\textbf{Dust Shear-Free Theorem} \cite{Ell67}: if a  dust solution
of the EFE (possibly with a cosmological constant) is shearfree in a
domain $U$, it cannot both expand and rotate in $U$: \be
\{\dot{u}^a=0,\,\, \sigma_{ab}= 0\} \Rightarrow \omega\Theta=0.
\l{theorem}\ee
\end{quote}
This is an exact result, obtained by utilizing all the field
equations. A covariant proof is given in \cite{SenSopSze98}.
Applying this theorem to the cosmological context, consider a
shear-free dust-filled universe that expands. Then (\ref{theorem})
shows $\omega_{ab} = 0$, so from the above equations
\begin{eqnarray}
 \{\sigma_{ab} =0, \,\, \Theta>0\} \Rightarrow E_{a b } = 0,\,\,H^{ab} = 0,
 \,\,\3nab^{a}\Th  =0.\label{e521}
\end{eqnarray}
The space-time is conformally flat and the universe is a
Friedmann-Lema\^{\i}tre-Robertson-Walker universe \cite{Ell71}.
These are thus the only expanding shear-free baryonic plus CDM
cosmological solutions, provided both these components move with the
same 4-velocity.

One should note that the result (\ref{e521}) does not require $\rho
> 0$. It is true in the vacuum case (with or without a cosmological
constant). Later generalizations considered perfect fluids rather
than pressure-free matter, so acceleration of the timelike
congruence was allowed; the result (\ref{theorem}) remains true in
all cases considered so far, indeed Collins conjectured \cite{Col86}
that all shear-free perfect fluids obeying a barotropic equation of
state must have either zero expansion or zero vorticity. Senovilla,
Sopuerta and Szekeres \cite{SenSopSze98} summarized results obtained
towards proving this conjecture, and gave a fully covariant proof
that shear free solutions with the acceleration vector proportional
to the vorticity vector (including the case of vanishing shear) must
be either non-expanding or non rotating. Van den Bergh
\cite{vanBer99} gave a tetrad-based approach for two particular
cases require a special treatment, namely $p + 1/3 \rho = $
constant, and $p - 1/9 \rho=$ constant, as well as the equation of
state $p = (\gamma-1)\rho + constant$. Van Den Bergh, Carminati, et
al \cite{vancarKar07,Caretal09} showed the result is generically
true for shear-free perfect fluid solutions of the Einstein field
equations where the fluid pressure satisfies a barotropic equation
of state and the spatial divergence of the magnetic part of the Weyl
tensor is zero.

Can one get models other than FLRW in these cases? Collins showed
\cite{Col85} that for irrotational shearfree perfect fluids obeying
a barotropic equation of state $p=p(\mu)$ and with nonzero
acceleration, one can get spherically symmetric Wyman solutions, or
models that are plane symmetric, and either spatially or temporally
homogeneous. In all cases, when the space-time is sufficiently
extended, the fluid exhibits unphysical properties. Consequently
shear-free expanding barotropic perfect fluids must either be FLRW,
or must be restricted to local regions where these conditions hold.
Thus it turns out that the FRW models are the only shear-free
barotropic perfect fluid  models in which the matter is physically
reasonable globally \cite{Col86}.

Overall, these results show clearly how restrictive the shear-free
result is for plausible fluid models. It will of course not be true
for ``imperfect fluids'' with arbitrary equations of state: one can
then just run the field equations from left to right to determine an
unphysical form of ``matter'' that will give the desired result.
Such calculations have no physical significance. One should note
here that despite what one might have thought at first, although
shear-free solutions are necessarily self-similar (they map the
orthogonal 3-spaces conformally onto each other), the converse is
not necessarily true: self similar solutions need not be shear-free
(see e.g. \cite{CarCol00}).

This is related to the idea of perfect fluid as follows: a continuum
description of matter is underlain by a kinetic theory description.
Now it follows from kinetic theory for a collision-free fluid that
if there is non-zero shear, there will be an anisotropic stress
(the non-zero shear will generate anisotropy in the particle distribution
function which will then result in an anisotropic pressure \cite{EllTreMat83}),
hence a perfect fluid description will not then
apply. This will also be true if there are collisions leading to
a non-zero shear viscosity, because of the relation
 \be \pi_{ab} = \lambda\, \sigma_{ab} \ee determining the anisotropic pressure $\pi_{ab}$, where
$\lambda$ is the shear viscosity coefficient \cite{Ehl61,Ell71}.
It follows that if one has a perfect fluid,

\be \{\pi_{ab} = 0, \lambda\neq0\} \Rightarrow  \{\sigma_{ab} =0\}. \ee
But kinetic theory shows that $\lambda \neq 0$ for realistic fluid
descriptions based on kinetic theory with collisions \cite{IsrSte76,EllTreMat83}.
Hence {\it an exact perfect fluid description of continuous matter
implies zero shear}, and the above results apply.

The moral is that realistic matter cannot be accurately represented
by a perfect fluid description. The difference may be small, but is
important in principle.

\subsection{The Newtonian limit}
This shearfree result is an exact result for the full non-linear
EFE: no approximations have been made, and is completely general: it
holds for any spatially homogeneous or inhomogeneous model. This
raises a very interesting situation as regards the Newtonian limit
of the EFE, because this result is not true in the Newtonian case.

The key point here is that Newtonian Gravity is not independent of
General Relativity: it derives from General Relativity in special
conditions. Specifically, it is a limiting form of General
Relativity, valid in particular circumstances (when matter relative
motion is at low speeds, and there are no gravi-magnetic effects or
gravitational waves, which will be true if the magnetic part of the
Weyl tensor is zero). Consequently the properties of Newtonian
gravity should follow from those of general
relativity.\footnote{This is in contrast to the view of some
astrophysicists that Newtonian theory is the true gravitational
theory, and GRT a set of small corrections to be made to this true
theory.}  One should note here that to obtain Newtonian cosmological
models, one has to use a description in terms of a gravitational potential where
one allows the potential to diverge at infinity, so it is not
strictly Newtonian theory, but rather an extension of the theory, where also
the idea of acceleration is generalized. With these generalizations, one
can obtain viable Newtonian cosmological models for the dynamical behavior
in a matter dominated era\footnote{This is not possible for a radiation dominated era.}
(the relevant equations and references are given in \cite{Ell71}).

 The major point then is that there are shearfree
solutions of the Newtonian equations for pressure-free matter in
cosmology that both expand and rotate; specific examples have been
given by Narlikar \cite{Nar63}. Consequently, the Newtonian limit is
singular. Consider a sequence $GRT (i)_{\sigma=0}$ of relativistic
shearfree dust solutions with a limiting solution $GRT
(0)_{\sigma=0}$ that constitutes the Newtonian limit of this
sequence. The latter solution will necessarily satisfy
(\ref{theorem}) because every solution $GRT (i)_{\sigma=0}$ in the
sequence does so. Newtonian solutions $NGT_{\sigma=0}$ that do not
satisfy (\ref{theorem}) are thus not obtainable as limits of any
sequence of relativistic solutions $GRT (j)_{\sigma=0}$. Assuming
Einstein's field equations represent the genuine theory of
gravitational interactions in the physical Universe, this result
tells us that not all Newtonian cosmological solutions are
acceptable approximations to the true theory of gravity.

An important application of this result is as follows: Narlikar has
shown \cite{Nar63} that shearfree and expanding Newtonian
cosmological solutions can have vorticity that spins up as the
universe decreases in size, and hence causes a `bounce' (the
associated centrifugal forces avoid a singularity). This would be a
counter-example to the cosmological singularity theorems of Hawking
and Penrose (see \cite{HawEll73}), if there were GRT analogues of
these singularity-free cosmological solutions; but the shearfree
result (\ref{theorem}) shows there are no such GR solutions. This is
a remarkable way in which the exact properties of GR dust solutions
support the results of the Hawking-Penrose singularity theorems,
obtained by completely different methods.

This is a case where the Newtonian models are very misleading. The
Newtonian limit is singular in such cases; so we need to be cautious
about that limit in other situations of astrophysical and
cosmological interest.

\subsection{The linearised case}
It is of considerable interest then whether the result
(\ref{theorem}) holds in the case of linearised perturbations of
FLRW universe models. It has recently been shown\footnote{A-M Nzoiki, R Goswami,
 P K S Dunsby, and G F R Ellis, in preparation.} that it holds in this case too: if a
perfect fluid with equation of state $ p = k\rho$ in an almost FLRW
universe is shear-free, then it must be either expansion-free or
rotation-free. Thus linearization does not lose this property. This
makes the situation even more remarkable: for then the linearised
solutions - almost universally used to study the formation of
structure by gravitational instability in the expanding universe,
and believed to result in standard local Newtonian theory - also
does not give the usual behaviour of Newtonian solutions in
interesting cases. The moral of the story is that you can't believe
a Newtonian solution unless it can be shown to be the limit of a
family of GRT solutions - else it may lead you badly astray (as in
the case of the shear-free expanding and rotating solutions found by
Narlikar).

\section{Null flows}
I now turn to the null case. The kinematic definitions of expansion,
shear and vorticity for congruences of null geodesics were given by
Ehlers and Sachs \cite{EhlSac61}, with a tetrad version 
being given by Newman and Penrose \cite{NewPen62} (summaries are
given in \cite{HawEll73}, \cite{PenRin86}). This proceeds in
parallel to the analysis for the timelike case, with two crucial
differences: the geodesic vector field $k^a = dx^a/dv$ is null
($k^ak_a = 0$) rather than timelike; and the projection is into a
two-dimensional spacelike screen space orthogonal to $k^a$ and an
observer $u^a$, instead of into a 3-dimensional space as in the case
of $u^a$ (so in this section, $<..>$ denotes trace-free
2-dimensional orthogonal projection to $k^a$ and $u^a$).

For an irrotational null geodesic congruence, the optical scalars
$\hat{\theta}$ (expansion, given by $\nabla_a k^a =2\hat{\theta}$)
and $\hat{\sigma}_{ab}= \hat{\sigma}_{<ab>}$ (shear) satisfy the
{\it Sachs equations}
\begin{eqnarray}
\frac{\dd\hat{\theta}}{\dd v} + \hat{\theta}^2 +
\vert\hat{\sigma}\vert^2 &=& \Phi_{00},\label{e125}
\\
\frac{\dd\hat{\sigma}_{ab}}{\dd v}  + 2\hat{\theta}\,
\hat{\sigma}_{ab} &=&  \Psi_{ab}
\label{e126}
 \end{eqnarray}
where the Ricci tensor term is $\Phi_{00}:=\kappa(\rho+p)$,
determined by the matter at each point,  and the Weyl tensor term is
$\Psi_{ab} := k^c C_{c<ab>d} k^d$, determined by matter elsewhere
plus boundary conditions.

  These are obviously in analogy to the timelike case
(\ref{e2}), (\ref{e5}) (there is no analogue to (\ref{e4}) because
the vector field $k^a$ is a gradient: $k_a = \nabla_a \phi$, and so
is irrotational). Thus one can again refer to {\em Ricci focusing}
caused pointwise by the matter distribution $\Phi_{00}$ inside the
beam,\footnote{It is noteworthy that the cosmological constant does
not enter here: it has no direct influence on null focusing.} and
{\em Weyl focusing} caused by the Weyl tensor term $\Psi_{ab}$
generated non-locally by matter outside the beam. It is the latter
effect that underlies gravitational lensing and consequent focussing
of the null geodesic rays. The full set of equations whereby these
non-local effects take place are given in terms of the spin
coefficient equations in (\cite{PenRin84}, pp.248-249); these are
equations (\ref{e2}) - (\ref{eq:divH}) above expressed in terms of a
null vector basis, plus equations determining the tetrad rotation
coefficients and giving the tetrad components relative to a
coordinate basis (required to get a complete set of equations).

These equations immediately imply \be \{\Phi_{00} = 0,
\hat{\theta}=0\} \Rightarrow \hat{\sigma} = 0\ee i.e. an irrotational
null congruence in empty space cannot shear without either expanding
or contracting. If we had included twist, the conclusion would have
been altered to, an irrotational null congruence in empty space
cannot shear without either expanding/contracting or twisting: a
kind of inverse of the timelike no-shear result.\footnote{There is
no corresponding timelike result if the cosmological constant is
non-zero.}

\subsection{The geometry of the Weyl tensor} The Petrov classification
describes the possible algebraic properties of the Weyl tensor at
each event in a Lorentzian manifold. It was first given in terms of
an orthonormal basis by Petrov \cite{Pet54},  and is nicely
described by Ehlers and Kundt \cite{EhlKun60} and by Penrose and
Rindler (\cite{PenRin86}, pp. 242-246).

The relation to null vectors was given by Ehlers and Sachs
\cite{EhlSac61}, showing how the timelike and spacelike Weyl
eigenbivectors are associated with preferred null vectors, called
the \emph{principal null directions} (PND's) of the Weyl tensor. The
condition for $k^a$ to be a principal null direction of the Weyl
tensor is \be k_{[e}C_{c]ba[d}k_{f]}k^c k^d = 0\label{pnd}. \ee In
general, there are four uniquely determined PND's at each point if
the Weyl tensor is non-zero, but in degenerate cases two or more
PND's may coincide; the degenerate PND's satisfy more restrictive
equations than (\ref{pnd}):  $k^a$ is a degenerate eigendirection of
the Weyl tensor iff \be C_{abc[d} k_{e]} k^b k^c = 0.
\label{dpnd}\ee The different possibilities lead to the six Petrov
types, succinctly described using a spinor formalism (\cite{Pen60};
\cite{PenRin86}, pp. 223-226); these different algebraic types
correspond to different physical situations. The Petrov types are,
\begin{itemize}
\item Type I  : four simple PNDs (generic: realistic models),
\item Type II : one double and two simple PNDs,
\item Type D  : two double principal null directions (massive objects with symmetry: e.g,
Schwarzschild),
\item Type III: one triple and one simple PND,
\item Type N : one quadruple PND: $C_{abcd}k^d = 0$ (pure gravitational waves:
e.g. a plane gravitational wave),
\item Type O: the Weyl tensor vanishes: $C_{abcd} = 0$ (no tidal forces: e.g. a FLRW
universe).
\end{itemize}
Type I is generic, all the others are algebraically special Weyl
types (they all have repeated PNDs) and so correspond to restricted
physical situations with symmetries.

\subsection{The Goldberg Sachs Theorem}
As in the timelike case, for essentially the same reasons, existence
of shear-free null congruences (discussed in \cite{PenRin86}, pp.
189-199)\footnote{Shear free null congruences play an important role
in twistor geometry (\cite{PenRin86}, Chapter 7).} places strong
restriction on the spacetime. We immediately find
\begin{eqnarray}
\frac{\dd\hat{\theta}}{\dd v} + \hat{\theta}^2
 &=& \Phi_{00},\label{e125a}
\\
0 &=&  \Psi_{ab}, \label{e126a}
 \end{eqnarray}
 the first showing that no new gravitational information can enter the congruence
 as it travels from the source to the observer (only the matter
 encountered by the null rays can cause convergence) and the second
 shows that shear-free null geodesics are PNDs (\cite{PenRin86}, (7.2.14) and (7.3.2)).
 In the vacuum case, further restrictions occur, captured in an important result by Joshua
Goldberg and Rainer Sachs:
\begin{quote}
\textbf{Goldberg-Sachs Theorem} \cite{GolSac62}: A vacuum metric
admits a shear free null geodesic congruence $k^a$ if and only if
$k^a$ is a degenerate eigendirection of the Weyl tensor (equation
(\ref{dpnd}) is true).
\end{quote}
This proves that a vacuum solution of the Einstein field equations
will admit a shear-free null geodesic congruence if and only if the
Weyl tensor is algebraically special. Shortly after the Goldberg and
Sachs paper, an alternative proof was given by Newman and Penrose
\cite{NewPen62}, using a tetrad or spinor formalism. A
generalisation was given by Robinson and Schild \cite{RobSch63},
establishing a connection between algebraic degeneracy of the Weyl
tensor, the existence of a null geodesic shear-free congruence, and
restrictions on the Ricci tensor which are weaker than the
requirement that there be empty space. in particular, they showed
that the gravitational field due to any Maxwell field with
shear-free rays is algebraically special. An even more generalized
version is given by Penrose and Rindler (\cite{PenRin86}, pp.
195-198).

The theorem is useful in searching for algebraically special vacuum
solutions, which is very helpful because almost all solutions we can
write down in exact form are algebraically special, corresponding to
restricted matter distributions and boundary conditions; examples
are the Kerr solution and plane gravitational waves.

\subsection{The News function}
Given the discussion above, one might expect that shear relates to
the way information is conveyed along bundles of null geodesics,
with restrictions on that information when the shear is zero. This
expectation seems to be borne out by the analysis of axisymmetric
vacuum spacetimes by Bondi, van den Berg and Metzner
\cite{BonVanMet62}, conveniently summarised in the book by D'Inverno
\cite{Din96}. On using a null coordinate system where the past null
cones of the central observer are given by $\{u = const\}$ where $u$
is the retarded time and $r$ is a measure of distance down the past
light cone, the metric is the Bondi metric \be ds^2= -(\frac{V}r
e^{2\beta} - r^2 e^{2\gamma} U^2) du^2 + 2 e^{2\beta} du dr + 2 U
r^2 e^{2\gamma} du d\theta - r^2 ( e^{2\gamma} d\theta^2 +
e^{-2\gamma} \sin^2\theta d\phi^2 ) \label{bondi}\ee where $V =
V(u,r,\theta)$, $U = U(u,r,\theta)$, $\beta = V(u,r,\theta)$,
$\gamma = \gamma(u,r,\theta)$. Then solving the vacuum EFE
asymptotically,
 $V = r - 2M + O(r^{-1}),$ 
$\beta = - n^2/4r^2 + O(r^{-3}),$ 
$4q_{,u} = 2Mn - d_{,\theta} + d \cot\theta,$ 
and \be \gamma = \frac{n(u,\theta)}{r} + \frac{q(u,\theta)}{r^3} +
O(r^{-4}). \ee
The mass of the system as measured at
infinity is the Bondi Mass \be m(u) = \frac{1}{2}\int
M(u,\theta)\sin\theta d\theta \ee
 The shear of the radially outgoing null rays is \be \hat{\sigma} =
n(u,\theta)/r^2 \label{shear}.\ee The initial data is
$\gamma(u,r,\theta)$ on an initial value null surface, plus
$n_{,u}(u,\theta)$ which determines the evolution of the source and
so is called the \emph{News Function}, determined by the first time
derivative of the shear. Finally \be m_{,0} = - \frac{1}{2}
\int_0^\pi (n_{,0})^2 \sin\theta d\theta\ee which is non-positive
and so shows that \emph{there is mass loss if and only if there is
news}.

 The Weyl tensor up to order $1/r$ has the
non-zero outgoing radiation component
\begin{eqnarray}
\Psi_0 &=& -n_{,uu}/r = - r \,\hat{\sigma}_{,uu} \label{Psi} 
\end{eqnarray}
so Bondi's formula says that the Bondi mass of the system decreases
if and only if there is outgoing radiation, and so if and only if
the second time derivative of the shear is non-zero. As stated in
the abstract, ``\emph{It is shown that the flow of information to
infinity is controlled by a single function of two variables called
the news function. Together with initial conditions specified on a
light cone, this function fully defines the behaviour of the
system..... The principal result of the paper is that the mass of a
system is constant if and only if there is no news; if there is
news, the mass decreases monotonically so long as it continues}.''
This confirms the key role played by shear in conveying news, and
hence in the mass loss carried out by outgoing gravitational
radiation. The result remains true if the assumption of axisymmetry
is dropped \cite{Sac62}.

Thus we have confirmation of the key concept of this paper: the
shear of geodesics, generated by the Weyl tensor, conveys
information about the gravitational field due to gravitating bodies.
If the shear is zero, this information is very limited and the space
time dynamics is highly constrained.

\subsection{The discontinuous limit}
However there is an intriguing context where this picture seems to
fail. The outgoing gravitational waves discussed in the previous
section should at large distances asymptotically become plane
gravitational waves.

The pp wave geometries \cite{BalAic07} are given in terms of null
coordinates by \be ds^2 = -2 du dv + h_{ij}(x,u)dx^idx^j \ee with a
covariantly constant null vector field $\xi^a = (\partial/\partial
u)^a$ which is consequently a Killing vector whose expansion and
shear vanish. Hence they are subset of null shear-free solutions;
the Weyl tensor is type N and this vector field is a 4-fold
degenerate PND (thus satisfying the Goldberg-Sachs theorem). Plane
gravitational waves  \cite{Bon57} are a special class of vacuum
pp-wave where \be ds^2 =
 - 2dudv + [a(u)(x^2 - y^2) + 2b(u) x y]du^2 + dx^2 + dy^2 \label{gravwave}\ee Here,
$a(u)$ and $b(u)$ can be any smooth functions; they specify the
amplitude of the two polarization modes of gravitational radiation.
The waves can convey arbitrary messages by the time variation of these
modes along successive light cones.

The issue then is the following:

 1: The Bondi news function for asymptotically flat vacuum metrics is the time derivative of the shear.
Hence, no shear implies no news.

2: The BMS metrics (\ref{bondi}) should become plane gravitational
waves asymptotically at infinity.

3: The pnd vector field of a plane gravitational wave is necessarily
an exactly shear-free geodesic congruence (Goldberg-Sachs).

4: But plane fronted gravitational waves can freely carry news: they
have two free polarisation functions, even though their shear is
exactly zero.

5: Hence the Bondi metrics should tend to a state where there is no
shear at infinity, hence no news, yet effective news transfer is
possible in the exact limiting spacetime.

How does this all fit together? Presumably it is a question of
relative orders of decay near infinity: but it is not obvious how it
is coherent! One might have thought that zero shear meant zero news:
but this is not the case!

There is of course a shear associated with plane waves, which is
essential for the existence of the waves. Null geodesics which
intersect the null hyperplane histories of the plane waves must have
shear (if they don't then the plane waves don't exist). Other
manifestations of this phenomenon are Penrose's observation that an
impulsive plane gravitational wave acts as an astigmatic lens and
also that colliding plane waves generate shear after collision. But
the paradox remains: the PND congruence is shear free, and indeed as
a consequence the geometry is very limited (it has a large symmetry
group) but still can convey arbitrary information.

\subsection{Linearised gravity}
There is no Newtonian analogue of the Goldberg-Sachs theorem, as
there are no equivalent concepts there. But one can consider the
linearized version of the result. It has been shown by Dain and
Moreschi \cite{DaiMor05} that a corresponding theorem will not hold
in linearized gravity, that is, given a solution of the linearised
Einstein field equations admitting a shear-free null congruence,
then this solution need not be algebraically special.

This is a warning of the perils of using linearised results for a
non-linear theory: some key results may not be valid in the
linearised case, even though they are an exact result of the
non-linear theory. Hence as in the case of the Newtonian limit of
the timelike case, here we can consider a sequence of exact
shearfree solutions that tend to a linearised shear free solution:
the limiting properties of the exact solutions differ from the
properties of the linearised solution. Hence for example special
solutions of the exact and linearised equations may have differen
properties.

\section{Conclusion}
Shear of geodesic curves plays a crucial role in conveying
information regarding the state of matter in a region, underlying
structure formation in the timelike case and gravitational lensing
in the null case.

I have compared the timelike and null cases of shear-free geodesics.
In each case the allowed exact solutions are strongly restricted,
albeit in rather different ways.  The argument makes clear that the
two contexts are related in the following way: one would not expect
gravitational radiation to be emitted by a shear-free dust flow, and
that is indeed the case (combine (\ref{e521}) and (\ref{radn})). We
have also seen that there are intriguing limiting questions in each
case: shear-free solutions don't have the same implications in
Newtonian theory (timelike case) and linearised gravity (the null
case). Pursuing these issues might be interesting.

One further line of of investigation that may also be interesting is
to pursue the stability of the Goldberg-Sachs result in the
following sense: an ``almost Birkhoff theorem'' has recently been
proven \cite{GosEll11}. This shows that if the conditions for
Birkhoff's theorem are almost true, then Birkhoff's theorem will be
approximately satisfied. An ``Almost Goldberg-Sachs theorem'' would
similarly show that if the conditions for the Goldberg-Sachs theorem
are almost true, then Goldberg-Sachs theorem will be approximately
satisfied. If this were not true, that might throw light on the
failure of the asymptotic limit discussed in the previous section.\\

It is a pleasure to dedicate this paper to Josh Goldberg, who has
done so much for GRG studies. I thank Peter Hogan for helpful
comments, and Roy Maartens for suggestions that have greatly
improved this paper.



\end{document}